\documentclass[11pt,letterpaper]{article}
\pdfoutput=1
\usepackage{jheppub}
\usepackage{amsfonts, amsthm}
\usepackage[english]{babel}
\usepackage[utf8]{inputenc}
\usepackage{slashed}
\hypersetup{unicode}

\newcommand{\ti}{\tilde}

\newcommand{\eq}{\begin{equation}}
\newcommand{\feq}{\end{equation}}
\newcommand{\eqn}{\begin{eqnarray}}
\newcommand{\feqn}{\end{eqnarray}}

\newcommand{\ma}[1]{\mbox{$\mathcal{#1}$}}

\newcommand{\D}{{\rm d}}

\title{On the integrability of Einstein-Maxwell- \\
(A)dS gravity in presence of Killing vectors}

\author{Dietmar Klemm, Masato Nozawa and Marco Rabbiosi}
\affiliation{Dipartimento di Fisica, Universit\`a di Milano, Via Celoria 16, 20133 Milano, Italy, \\
and \\
INFN, Sezione di Milano, Via Celoria 16, 20133 Milano, Italy.
}
\emailAdd{dietmar.klemm@mi.infn.it}
\emailAdd{masato.nozawa@mi.infn.it}
\emailAdd{marco.rabbiosi@mi.infn.it}
\preprint{IFUM-1039-FT}

\abstract{We study some symmetry and integrability properties of four-dimensional Einstein-Maxwell
gravity with nonvanishing cosmological constant in the presence of Killing vectors. First of all, we consider
stationary spacetimes, which lead, after a timelike Kaluza-Klein reduction followed by a dualization
of the two vector fields, to a three-dimensional nonlinear sigma model coupled to gravity, whose
target space is a noncompact version of $\mathbb{C}\text{P}^2$ with $\text{SU}(2,1)$ isometry group. It is shown that the potential for the scalars, that arises from the cosmological constant in four dimensions,
breaks three of the eight $\text{SU}(2,1)$ symmetries, corresponding to the generalized Ehlers and the two Harrison transformations. This leaves a semidirect product of a one-dimensional Heisenberg group
and a translation group $\mathbb{R}^2$ as residual symmetry. We show that, under the additional
assumptions that the three-dimensional manifold is conformal to a product space
$\mathbb{R}\times\Sigma$, and all fields depend only on the coordinate along $\mathbb{R}$,
the equations of motion are integrable. This generalizes the results of Leigh et al.~in
arXiv:1403.6511 to the case where also electromagnetic fields are present. 
In the second part of the paper we consider the purely gravitational spacetime admitting a second Killing vector that commutes with the timelike one. We write down the resulting two-dimensional action and discuss its symmetries. If the fields depend only on one of the two coordinates, the equations of motion are again integrable, and the solution turns out to be one constructed by Krasi\'nski many years ago.
}


\begin{document}
\maketitle
\flushbottom

\section{Introduction}

Exact solutions to Einstein's field equations and to their supergravity generalizations have been playing
an important role in many developments of general relativity, string theory and high energy physics~\cite{Stephani:2003tm}. For instance they can teach us a lot of insight into the theoretical aspects of general relativity such as  the vacuum structure, uniqueness theorems and so on.  
The importance of exact solutions, though, is not limited to the classical situation, but extends also
to quantum gravity. Indeed, much of our current knowledge on quantum effects in strong gravitational
fields comes from the study of classical black hole solutions. 

Generically, the construction of exact solutions to general relativity is a notoriously difficult
problem, since the underlying field equations are a system of coupled nonlinear partial differential equations  of second order. Nevertheless, one may hope that this system becomes integrable when some sufficient amount of symmetry is present. During late 1970s, a variety of independent groups have established the integrability  of Einstein's vacuum equations in stationary and axisymmetric systems. 
This includes the discovery of B\"acklund transformations by Harrison~\cite{Harrison} and by 
Neugebauer~\cite{Neugebauer:1979iw}, and a Lax-pair representation by Belinsky and
Zakharov (BZ) \cite{Belinsky:1971nt}\footnote{Shortly before \cite{Belinsky:1971nt} appeared,
Maison \cite{Maison:1978es} was able to rewrite the stationary axisymmetric vacuum Einstein equations
as a `linear eigenvalue problem in the spirit of Lax', and noticed that this could `nourish some hope
that a method similar to the inverse scattering method may be developed'.}.
 In particular, the results of \cite{Belinsky:1971nt} have been generalized in many directions, e.g.~to the Einstein-Maxwell system \cite{Alekseev:1980,Alekseev:2009zz}, five-dimensional general relativity \cite{Pomeransky:2005sj} and five-dimensional minimal ungauged supergravity \cite{Figueras:2009mc} (cf.~also \cite{Bouchareb:2007ax,Clement:2007qy,Compere:2009zh}). These techniques of integrable
systems allow us to construct nontrivial new solutions starting from a given seed by adding solitons in a simple algebraic manner. Perhaps one of the most exciting recent achievements for the applications of these techniques is the inverse scattering construction of black objects admitting multiple horizons with various topologies in five-dimensional pure gravity (see e.g, \cite{Iguchi:2011qi} for a comprehensive review).

Apart from the construction of numerous exact solutions, the underlying mathematical structure behind these integrable systems has been worked out by many authors.  Prior to the studies of integrability, 
Geroch made a pioneering analysis of solution-generating methods for Ricci flat spaces in presence of
a single \cite{Geroch:1970nt} and two mutually-commuting Killing vectors \cite{Geroch:1972yt}
(see \cite{Kinnersley:1977pg,Kinnersley:1977ph,Kinnersley:1978pz,HKX} for an electromagnetic generalization). In the presence of two commuting Killing vectors, the target space isometry does not commute with the internal ${\rm SL}(2,\mathbb R)$ symmetry, giving rise to an infinite affine Lie group
called the Geroch group. Hauser and Ernst were able to prove the conjecture that any stationary and axisymmetric solution can be derived in principle from a Minkowski seed by the Geroch group~\cite{HauserErnst}. Cosgrove addressed the interrelationships between several solitonic systems and the Geroch group~\cite{Cosgrove:1980fx}.  Later on, Breitenlohner and Maison (BM) unraveled the group-theoretical structure of solitonic methods from the standpoint of a Riemann-Hilbert problem~\cite{Breitenlohner:1986um}. A very nice general analysis of the relation between the BM group structure and the inverse scattering method of the BZ approach was recently given in \cite{Katsimpouri:2012ky}. In addition, a close connection to nonlinear sigma models has also been widely discussed~\cite{Breitenlohner:1998cv}. 

In view of possible AdS/CFT (and many other) applications, one may thus ask whether similar integrability
properties still hold in the presence of a cosmological constant, for example for stationary axisymmetric
Einstein spaces in four dimensions. The introduction of a negative cosmological constant has a strong impact upon the spectrum of black holes.  Of most prominent is that the horizon of 
an asymptotically AdS black hole can be a compact Riemann surface of any genus \cite{Lemos:1994xp,
Mann:1996gj,Vanzo:1997gw,Cai:1996eg}. This is in contrast to black holes in asymptotically flat spacetimes \cite{Hawking:1971vc,Hawking:1973uf}. This is not the end of the story, since even more exotic possibilities exist, e.g.~noncompact horizons with finite area \cite{Gnecchi:2013mja,Klemm:2014rda} (for generalizations to higher dimensions and further discussions of the physics of these solutions cf.~also \cite{Hennigar:2014cfa,Hennigar:2015cja}).
One may thus expect a rich spectrum of black objects in presence of a cosmological constant, with many
of them perhaps still to be discovered. It is clear that the integrability of stationary axisymmetric Einstein
spaces would simplify enormously the construction of such solutions. A main obstruction for this program is that the metric cannot be cast into the Weyl-Papapetrou form in the presence of a cosmological constant. It is therefore obvious that the techniques available in the absence of $\Lambda$ cannot be straightforwardly applied. 

First steps in the investigation of the integrability properties with nonvanishing $\Lambda$ were
undertaken in \cite{Beppe,Charmousis:2006fx,Astorino:2012zm,Leigh:2014dja}. These papers developed
solution-generating techniques for the stationary vacuum \cite {Charmousis:2006fx,Leigh:2014dja}
and electrovac \cite{Astorino:2012zm} Einstein equations with a cosmological constant, in
four \cite{Astorino:2012zm,Leigh:2014dja} and higher \cite{Charmousis:2006fx} dimensions. 
The cosmological constant leads to a potential in the dimensionally reduced system, breaking the
symmetries of the original sigma model, and thus the usual solution-generating techniques can't be
applied anymore. Still, some restricted formalism of solution-generating method is still applicable. 
In spite of this limited utility, they turn out indeed fruitful to generate some new exact solutions~\cite{Astorino:2012zm}.  

Here we shall make a first step towards a systematic investigation of the integrability properties
of Einstein-Maxwell gravity with nonvanishing cosmological constant in four dimensions, 
by extending the work of \cite{Leigh:2014dja}. In the first part of this paper we consider
stationary spacetimes which are described, after a timelike Kaluza-Klein reduction followed by a dualization of the two vector fields, by a three-dimensional nonlinear sigma model coupled to gravity, whose target space admits an $\text{SU}(2,1)$ isometry group. 
It is shown that the potential for the scalars, that arises from the cosmological constant in four dimensions,
breaks three of the eight $\text{SU}(2,1)$ symmetries, namely the generalized Ehlers and the two
Harrison transformations. This leaves a semidirect product of a one-dimensional Heisenberg group
and a translation group $\mathbb{R}^2$ as residual symmetry. We show that, under the additional
assumptions that the three-dimensional manifold is conformal to a product space
$\mathbb{R}\times\Sigma$, and all fields depend only on the coordinate along $\mathbb{R}$,
the equations of motion are integrable.  Subsequently, we consider the purely gravitational case and assume the existence of a second Killing vector that commutes with the timelike one, i.e., we focus on stationary and axisymmetric Einstein spaces. We write down the resulting
two-dimensional action and discuss its symmetries. If the fields depend only on one of the two
coordinates, the equations of motion are again integrable, and the solution turns out to be one
constructed by Krasi\'nski many years ago \cite{Krasinski:1974zza,Krasinski:1900zza,Krasinski:1994zz}.

The remainder of this paper is organized as follows: In the next section, 
we discuss the integrability of the stationary Einstein-Maxwell-$\Lambda$ system by assuming that the
base space takes a product structure $\mathbb{R}\times\Sigma$ and  the target space variables depend
only on a single coordinate. We derive the Reissner-Nordstr\"om-Taub-NUT-(A)dS metric by exploiting the Hamilton-Jacobi method.  This generalizes the results of \cite{Leigh:2014dja}
to the case where also electromagnetic fields are present. In section \ref{sec:Einst-Lambda}, 
we address the integrability of the Einstein-$\Lambda$ system, by assuming a second independent Killing vector, and derive an interesting class of solutions. Finally, we conclude in section \ref{sec:final-rem} with some final remarks. An appendix provides an attempt of a higher-dimensional generalization.

\section{Einstein-Maxwell-$\Lambda$ system}

In this paper, we focus on $3+1$-dimensional Einstein-Maxwell-(A)dS
gravity, with action\footnote{We use the signature $(-,+,+,+)$. The Ricci tensor is defined as
$R_{\mu\nu}=R^{\sigma}_{\;\mu\sigma\nu}=\partial_{\sigma}\Gamma_{\;\mu\nu}^{\sigma}-\partial_{\nu}\Gamma_{\;\mu\sigma}^{\sigma}+\Gamma_{\;\mu\nu}^{\rho}\Gamma_{\;\sigma\rho}^{\sigma}-\Gamma_{\;\mu\sigma}^{\rho}\Gamma_{\;\nu\rho}^{\sigma}$.} 
\begin{equation}
S=\frac 1{16\pi G} \int\D^{4}x\sqrt{-g}\left(R-F_{\mu\nu}F^{\mu\nu}-2\Lambda\right)\,, \label{eq:EML action}
\end{equation}
and equations of motion
\begin{equation}
R_{\mu\nu}-\frac{1}{2}Rg_{\mu\nu}+\Lambda
g_{\mu\nu}=2\left(F_{\mu\sigma}F_{\nu}^{\:\sigma}-\frac{1}{4}g_{\mu\nu}F_{\sigma\rho}F^{\sigma\rho}\right)\,, \qquad
\nabla_{\mu}F^{\mu\nu} = 0\,.
\end{equation}
The Faraday tensor can be locally expressed in terms of a gauge potential as
$F=\D A$. 

We shall investigate the integrability properties of the stationary Einstein-Maxwell-$\Lambda$ system,
which is an extension of the work in \cite{Leigh:2014dja}.

\subsection{Dimensional reduction}

Let us consider stationary spacetimes admitting a Killing field which is timelike at infinity.
Applying the algorithm of Kaluza-Klein reduction along the timelike direction, 
the metic and the gauge field can be decomposed as   
\begin{align}
\label{}
\D s^2=-e^{-\phi }(\D t+K_\alpha \D x^\alpha)^2+e^\phi h_{\alpha\beta}\D x^\alpha \D x^\beta \,, 
\qquad 
A= B(\D t+K_\alpha \D x^\alpha)+B_\alpha \D x^\alpha \,, 
\end{align}
where early greek indices refer to three dimensions, and the fields $h_{\alpha\beta}$, $K_{\alpha}$,
$B_{\alpha}$, $\phi$ and $B$ are $t$-independent. Here and in what follows, the indices $\alpha, \beta,..$ are raised and lowered by $h_{\alpha\beta}$ and its inverse. 
Then the effective three-dimensional Lagrangian derived from (\ref{eq:EML action}) becomes 
\begin{eqnarray}
\mathcal{L}^{(3)} &=& \sqrt{h}\left[R^{(3)}-\frac{1}{2}\partial_{\alpha}\phi\partial^{\alpha}\phi+\frac{1}{4}e^{-2\phi}K_{\alpha\beta}K^{\alpha\beta} + 2e^{\phi}\partial_{\alpha}B\partial^{\alpha}B\right.
\nonumber \\
&& \left.\quad-e^{-\phi}(G_{\alpha\beta}+K_{\alpha\beta}B)(G^{\alpha\beta} + K^{\alpha\beta}B) -
2\Lambda e^{\phi}\right]\,, \label{eq:Lagrangianreduced}
\end{eqnarray}
where $G_{\alpha\beta}\equiv \partial_{\alpha}B_{\beta}-\partial_{\beta}B_{\alpha}$ and
$K_{\alpha\beta}\equiv \partial_{\alpha}K_{\beta}-\partial_{\beta}K_{\alpha}$. It is convenient to dualize
the two vector fields to scalars, which can be implemented by adding to (\ref{eq:Lagrangianreduced}) a
piece containing two Lagrange multipliers $C$ and $\tilde\psi$ that ensure the Bianchi identities,
\begin{equation}
\tilde{\mathcal L}^{\left(3\right)} = \mathcal{L}^{(3)}+2C\epsilon^{\alpha\beta\gamma}\partial_{\alpha}
G_{\beta\gamma} + (\tilde{\psi}+CB)\epsilon^{\alpha\beta\gamma}\partial_{\alpha}K_{\beta\gamma}\,.
\label{Ltilde}
\end{equation}
Variation of \eqref{Ltilde} w.r.t.~$K_{\alpha\beta}$ and $G_{\alpha\beta}$ yields
\begin{equation}
K^{\alpha\beta} = \frac{2}{\sqrt{h}}e^{2\phi}\epsilon^{\alpha\beta\gamma}\omega_{\gamma}\,,
\qquad 
\omega_{\gamma} \equiv \partial_{\gamma}\tilde{\psi}+C\partial_{\gamma}B-
B\partial_{\gamma}C\,, \label{eq:dual1}
\end{equation}
and
\begin{equation}
G^{\alpha\beta}+K^{\alpha\beta}B=-\frac1{\sqrt{h}}e^{\phi}\epsilon^{\alpha\beta\gamma}
\partial_{\gamma}C\,. 
\label{eq:dual2}
\end{equation}
These equations express the field strengths in terms of the twist potential $\tilde\psi$ and the magnetic potential
$C$. Plugging (\ref{eq:dual1}) and (\ref{eq:dual2}) back into \eqref{Ltilde} leads (after dropping a tilde
on $\tilde{\cal L}^{(3)}$) to
\begin{equation}
\mathcal{L}^{(3)}=\sqrt{h}\left[R^{(3)}-\left\langle
J_{\alpha},J^{\alpha}\right\rangle -2\Lambda e^{\phi}\right],\label{eq:Lagrange3d}
\end{equation}
where  we have introduced the notation
\begin{align}
\left\langle J_{\alpha},J_{\beta }\right\rangle
\equiv \frac12\left[\partial_{\alpha}\phi\partial_{\beta}\phi+4e^{2\phi}\omega_{\alpha}\omega_{\beta}-
4e^{\phi}\left(\partial_{\alpha}B\partial_{\beta}B+\partial_{\alpha}C\partial_{\beta}C\right)\right] \,. 
\end{align} 
The equations of motion following from the Lagrangian (\ref{eq:Lagrange3d}) are the
three-dimensional Einstein equations
\begin{equation}
G_{\alpha\beta}^{(3)}+\Lambda e^{\phi}h_{\alpha\beta}=\left\langle J_{\alpha}, J_{\beta}
\right\rangle -\frac12{h}_{\alpha\beta}\left\langle J_\gamma, J^\gamma\right\rangle\,, \label{Einst-3d}
\end{equation}
supplemented by the divergence-type equations of motion
\begin{equation}
\nabla_{\alpha}[\partial^{\alpha}\phi+2e^\phi(B\partial^{\alpha}B+C\partial^{\alpha}C)-4e^{2\phi}\tilde{\psi}
\omega^{\alpha}]=2\Lambda e^\phi\,,\qquad\nabla_{\alpha}(e^{2\phi}\omega^{\alpha}) = 0\,,
\end{equation}
\begin{equation}
\nabla_{\alpha}(e^{\phi}\partial^{\alpha}B-2e^{2\phi}C\omega^{\alpha}) = 0\,,\qquad\nabla_\alpha(e^\phi\partial^{\alpha}C+2e^{2\phi}B\omega^\alpha)=0\,.
\end{equation}
\eqref{eq:Lagrange3d} describes a nonlinear $\sigma$-model with pseudo-Riemannian
target space coupled to Euclidean gravity in $d=3$, with
a potential. The latter breaks part of the target space isometries.

\subsection{Nonlinear $\sigma$-model and broken symmetries\label{sub:Nonlinear--Model}}

The target space $\Phi$ of the scalars in \eqref{eq:Lagrange3d} is a Bergmann space corresponding to  a noncompact version of
$\mathbb{C}\text{P}^2$ \cite{Breitenlohner:1987dg,Breitenlohner:1998cv,Astorino:2012zm}, namely 
it describes a coset space $\text{SU}(2,1)/{\text{S}(\text{U}(1,1)\times\text{U}(1))}$, 
endowed with the metric
\begin{equation}
\D s_{\Phi}^{2} =\ma G_{IJ}(\varphi)\D \varphi^I \D \varphi^J= \D\phi^{2}+4e^{2\phi}(\D\tilde{\psi}+C\D B-B\D C)^2-4e^{\phi}(\D B^2+\D C^2)\,, 
\label{target}
\end{equation}
where $\varphi^I=(\phi,\tilde \psi, B, C)$. 
One can easily verify that
\begin{align}
\label{target_prop}
R_{IJ} =-\frac 32 \ma G_{IJ}\,, \qquad 
C_{IJKL}=-\frac 12 \epsilon_{IJMN}C^{MN}{}_{KL} \,, \qquad 
D_I R_{JKLM}=0 \,. 
\end{align}
Here $R_{IJKL}$ and $C_{IJKL}$ are the Riemann and Weyl tensors constructed from the target space
metric $\ma G_{IJ}$ and the covariant derivative $D_I$. 
The Bergmann space is a special K\"ahler-Einstein manifold with negative curvature. 
The last equation of (\ref{target_prop}) is a differential characterization of a symmetric space, while 
the second equation implies a quaternionic structure~\cite{Hoegner:2012sq}. 

The eight Killing vectors of $\Phi$ generating the isometry algebra $\mathfrak{su}(2,1)$ are given by
\begin{eqnarray}
&&\xi_1 =\partial_{\tilde{\psi}}\,, \qquad \xi_2 = C\partial_{\tilde{\psi}}+\partial_B\,, \qquad
\xi_3 = -B\partial_{\tilde{\psi}}+\partial_{C}\,, \nonumber \\
&&\xi_4 = -C\partial_{B}+B\partial_{C}\,, \qquad \xi_5 =
-2\partial_{\phi}+2\tilde{\psi}\partial_{\tilde{\psi}}+B\partial_B+C\partial_C\,, \nonumber \\
&&\xi_6 = 4\tilde{\psi}\partial_{\phi}+\left[\frac12(e^{-\phi}-(B^2+C^2))^{2}-2\tilde{\psi}^{2}\right]
\partial_{\tilde{\psi}} \nonumber \\
&&\qquad +\left[C(e^{-\phi}-(B^2+C^2))-2\tilde{\psi}B\right]\partial_B-\left[B(e^{-\phi}-
(B^2+C^2))+2\tilde{\psi}C\right]\partial_C\,, \nonumber \\
&&\xi_7 = -4B\partial_\phi+\left[2\tilde{\psi}B-C(e^{-\phi}-(B^2+C^2))\right]\partial_{\tilde{\psi}} \nonumber \\
&&\qquad + (e^{-\phi}+B^2-3C^2)\partial_B + (4BC-2\tilde{\psi})\partial_C\,, \nonumber \\
&&\xi_8 = -4C\partial_\phi+\left[2\tilde{\psi}C+B(e^{-\phi}-(B^2+C^2))\right]\partial_{\tilde{\psi}} \nonumber \\
&&\qquad + (4BC+2\tilde{\psi})\partial_B+(e^{-\phi}+C^2-3B^2)\partial_C\,.
\end{eqnarray}
The first five Killing vectors represent infinitesimal transformations
that are linear in the scalars and comprehend a twist transformation,
two electromagnetic gauge transformations, an internal $\text{U}(1)$ transformation and
a scaling one. The remaining three are the most interesting, due to the nonlinearity
in the fields, and they are usually called generalized Ehlers transformation ($\xi_6$)
\cite{Ehlers:1957zz} and two Harrison transformations ($\xi_7,\xi_8$) \cite{Harrison:1980xxx}.

In order to see that these Killing vectors indeed generate the ${\rm SU}(2,1)$ symmetry, 
let us define
\begin{align}
E_2{}^1&=-\frac 14 [\xi_7+i \xi_8+i (\xi_3-i\xi_2)] \,, \qquad 
E_2{}^3=-\frac 14 [-(\xi_7+i \xi_8)+i (\xi_3-i\xi_2)] \,, \notag \\
E_1{}^3&=\frac 14(2 \xi_5+i \xi_1+2i \xi_6)\,, \quad 
E_1{}^1=H_1+E_3{}^3\,, \qquad 
E_2{}^2=H_2+E_3{}^3\,,\\
E_3{}^3&=-\frac 13 (H_1+H_2) \,, \qquad 
E_1{}^2=-(E_2{}^1)^* \,, \qquad E_3{}^1=(E_1{}^3)^*\,, \qquad 
E_3{}^2=(E_2{}^3)^* \,, \notag 
\end{align}
where $H_1$, $H_2$ are Cartan generators defined by
\begin{align}
H_1=\frac i2 \xi_1-i \xi_6 \,, \qquad 
H_2=\frac i4(\xi_1-6\xi_4-2\xi_6)\,,\qquad 
[H_1,H_2]=0 \,. \notag 
\end{align}
One can easily verify that these vectors $E_i{}^j$ ($i,j=1,2,3$) satisfy the $\mathfrak{su}(2,1)$ algebra
\begin{align}
\label{}
[E_i{}^j, E_k{}^l]=\delta_k{}^j E_i{}^l-\delta_i{}^l E_k{}^j \,, \qquad 
E_i{}^i=0\,.
\end{align}

Note that the dependence of the scalar potential
\begin{equation}
V(\phi) = -2\Lambda e^{\phi}
\end{equation}
on the dilaton $\phi$ breaks the invariance under nonlinear isometries and scalings. It is easy to see that
the latter is recovered if we admit a rescaling of $\Lambda$.
The theory described by (\ref{eq:Lagrange3d}) is thus invariant only under
$\text{SU}(2,1)/\text{H}_1$, where $\text{H}_1\subset\text{SU}(2,1)$ is a subgroup 
generated by $\xi_6$, $\xi_7$, $\xi_8$ corresponding to the Heisenberg algebra
\begin{equation}
[\xi_7,\xi_8] = 4\xi_6\,, \qquad [\xi_7,\xi_6] = [\xi_8,\xi_6] = 0\,.
\end{equation}
The five unbroken generators close themselves to form another one-dimensional Heisenberg subalgebra in semidirect sum with $\mathbb{R}^2$,
\begin{align}
\label{Killing_com2}
[\xi_2,\xi_3] &= -2\xi_1\,, \qquad [\xi_2,\xi_1] = [\xi_3,\xi_1] = 0\,, \nonumber \\
[\xi_i,\xi_4] &= {(\sigma_4)}_{i}^{\,j}\xi_j\,, \qquad [\xi_i,\xi_5] = {(\sigma_5)}_{i}^{\,j}\xi_j\,,
\end{align}
where $i,j=1,2,3$ and
\[
\sigma_4=\left(\begin{array}{ccc}
0 & 0 & 0\\
0 & 0 & 1\\
0 & -1 & 0\\
\end{array}\right)\,, \qquad 
\sigma_5=\left(\begin{array}{ccc}
2 & 0 & 0\\
0 & 1 & 0\\
0 & 0 & 1\\
\end{array}\right)\,.
\]
The Heisenberg algebra (\ref{Killing_com2}) realizes the fact that the constant $\phi$ space constitutes a
Nil manifold, viz, one can view the four-dimensional metric (\ref{target}) as a Wick-rotated Bianchi-II
universe.

The well-known solution-generating techniques \cite{Clement:1999bv,Clement:1997tx,Katsimpouri:2012ky}
based on group theory can thus not be applied in presence of a cosmological constant. Moreover, the
broken symmetries are also a first sign of the loss of complete integrability, valid for $\Lambda=0$
after another dimensional reduction \cite{Breitenlohner:1986um,Nicolai:2001xxx,Nicolai:1991tt}.
This implies also the inapplicability of the inverse scattering
method \cite{Belinsky:1971nt,Katsimpouri:2012ky}. In what follows, we shall perform an analysis
of some remaining integrability properties, extending the results of \cite{Leigh:2014dja}.

\subsection{Hamiltonian formalism and first integrals}

In the spacetime admitting a single Killing field, the sigma model still couples to the base space $h_{\alpha\beta} $ represented by three-dimensional Einstein gravity according to (\ref{Einst-3d}).
Because of the intricacy of this system, we usually simplify the problem by assuming further symmetries. In the absence of $\Lambda$, the base space is decoupled from the sigma model by assuming an axial Killing
field. More precisely, the metric without $\Lambda$ can be cast into the Weyl-Papapetrou form, and the base space part can be obtained by quadrature once the sigma model on $\mathbb R^2$ is solved. Unfortunately, this decoupling does not occur in the presence of $\Lambda$ as we will see in section~\ref{sec:Einst-Lambda}. 

In this section, we follow a different path to arrive at an integrable system. 
Along the lines of the argument in \cite{Leigh:2014dja}, we consider the case in which the base space admits only a single degree of freedom. 
Now we suppose that ${h}_{\alpha\beta}$ describes a warped product space $\mathbb{R}\times\Sigma$, with $\Sigma$ a two-dimensional manifold. Moreover we assume that all the scalar fields
depend only on the coordinate representing $\mathbb R$. To capture this more conveniently, let us introduce another scalar field $k$ that describes a rescaling of the three-dimensional metric $h_{\alpha\beta}$,
\begin{equation}
h_{\alpha\beta}=k\hat{h}_{\alpha\beta}\,.
\end{equation}
Absorbing the warp factor into $k$, 
$\hat h_{\alpha\beta}$ can be taken to be an unwarped product, 
\begin{equation}
\hat h_{\alpha\beta}\D x^\alpha \D x^\beta = \D\sigma^2 + \D\Omega^2\,, 
\label{eq:subspace}
\end{equation}
where $\D\Omega^2$ is the line element on $\Sigma$.  Under these settings, every quantity depends only on a single valuable $\sigma$. In this case the trace and the $\sigma\sigma$-component of the Einstein
equations \eqref{Einst-3d} become respectively
\begin{eqnarray}
\hat{R}^{(3)} &=& \frac1{2k^2}\left(\frac{\D k}{\D \sigma}\right)^2 - \langle J_{\sigma},J_{\sigma}\rangle
+ 2\Lambda ke^{\phi}\,, \label{eq:einsteintransverse} \\
\frac 1k \left(\frac{\D^2 k}{\D \sigma^2}\right) &=& \frac1{k^2}\left(\frac{\D k}{\D \sigma}\right)^2
-\langle J_{\sigma},J_{\sigma}\rangle -2\Lambda k e^\phi \,. 
\end{eqnarray}
It is clear that the scalar curvature $\hat{R}^{(3)}$ must be constant as a consequence of the fact that
the r.h.s.~of (\ref{eq:einsteintransverse}) depends only on $\sigma$ and the l.h.s.~is independent of
$\sigma$. Without further resrictions we can thus take $\hat{R}^{(3)}=2l$ with $l=0,\pm 1$,
so that $\Sigma$ must be a maximally symmetric space,
$\D\Omega_l^2=\D\theta^2+f_l^2(\theta)\D\varphi^2$, where
\begin{equation}
f_l(\theta) = \frac 1{\sqrt l} \sin (\sqrt l \theta)= 
\left\{\begin{array}{c@{\quad}l} \sin\theta\,, & l=1\,, \\
                                               \theta\,, & l=0\,, \\
                                               \sinh\theta\,, & l=-1\,. \end{array}\right.
\end{equation}
One obtains then a classical dynamical system with five degrees of freedom, with action
\begin{equation}
S = \int \D \sigma k^{\frac12}\left[\frac1{2k^{2}}\left(\frac{\D k}{\D \sigma}\right)^2 - \langle
J_{\sigma},J_{\sigma}\rangle +2l-2\Lambda ke^\phi\right]\,. \label{eq:action-one-dim}
\end{equation}
For future convenience we introduce a new evolution parameter $\tau$ defined by 
\[
k^{\frac32}e^\phi\D\sigma = \D\tau\,.
\]
With the new potential $\hat{V}=2\Lambda-\frac{2l}k e^{-\phi}$ and 
$\omega\equiv\omega_\tau$, 
the action (\ref{eq:action-one-dim}) can be expressed as $S=\int L \D \tau$ with a 
Lagrangian 
\begin{equation}
L = \frac12\left[e^{\phi}{k'}^2 - k^2 e^\phi{\phi'}^2 - 4k^2 e^{3\phi}\omega^2 + 4e^{2\phi}k^2
({B'}^2 + {C'}^2)\right] - \hat{V}\,, \label{eq:Lagrangianacap}
\end{equation}
where a prime denotes a derivative w.r.t.~$\tau$. It is easy to see that
(\ref{eq:einsteintransverse}) is the constraint $H\equiv L+2\hat V=0$.
It then turns out more convenient to pass to a Hamiltonian formulation rather than working in a 
Lagrangian description.  
After a Legendre transformation one gets
\begin{eqnarray}
H &=& \frac12\left[e^{-\phi}p_{k}^{2}-\frac{e^{-\phi}}{k^{2}}p_{\phi}^{2}-\frac{e^{-3\phi}}{4k^{2}}p_{\tilde{\psi}}^2\right. \label{eq:Hamiltonian} \\
&+&\left.\frac{e^{-2\phi}}{4k^2}\left(p_B^2 + p_C^2 - 2Cp_B p_{\tilde{\psi}}+2B p_C p_{\tilde{\psi}}+(B^2+C^2)p_{\tilde{\psi}}^2\right)\right]+\hat{V}\,. \nonumber
\end{eqnarray}
The solution of this dynamical system is highly linked to the existence of commuting constants of
motion. The Killing vector fields of $\mathfrak{su}(2,1)$ can be promoted to functions in phase space,
realizing a Lie algebra isomorphism, by means of the substitutions\footnote{Our convention for the
Poisson bracket is $\{A, B \}\equiv \Omega^{MN}\partial_MA \partial_N B 
=\sum_I \left(\frac{\partial A}{\partial q^I}\frac{\partial B}{\partial p_I}
-\frac{\partial A}{\partial p_I}\frac{\partial B}{\partial q^I}\right)$,  
where $\Omega=i\sigma_2$ is the symplectic form.}
\begin{equation}
\partial_{\varphi^I}\mapsto p_{\varphi^I}\,, \qquad [\cdot,\cdot]\mapsto\{\cdot,\cdot\}_{\text{PB}}\,,
\qquad \xi_i\mapsto -C_i\,,
\end{equation}
where $\{\varphi^I\}=\{\phi,\tilde\psi,B,C\}$ and $i=1,\ldots,8$. 
The minus sign in front of $C_i $ reflects the fact that the infinitesimal generators and the corresponding charges obey the same algebra up to the sign of the structure constants\footnote{This can be shown as
follows. Let $Q_i=Q_i(q^I, p_I)$ be first integrals obeying the Lie algebra
$\{Q_i, Q_j\}=f^k{}_{ij}Q_k$ and let us denote 
the corresponding Hamiltonian vector fields by 
$V_i^M=\Omega^{MN}\partial_NQ_i$. 
For any function $F=F(q^I,p_I)$ in phase space,  
we have a formula $V_{i}^M\partial_MF=-\{Q_i, F\}$. 
It follows that for the vector field 
$V_{ij}^M=\Omega^{MN}\partial_N\{Q_i,Q_j\}=f^k{}_{ij}V^M_k$, we obtain 
$V_{ij}^M\partial_MF=-\{\{Q_i,Q_j\}, F\}=-\{Q_i, \{Q_j, F\}\}+\{Q_j, \{Q_i, F\}\}=-[V_i, V_j]^M\partial_M F$, where at the second equality we used the Jacobi identity. 
This establishes $[V_i, V_j]=-f^k{}_{ij}V_k$, as desired.}.
The only nonvanishing Poisson
brackets between the $C_i$ and the Hamiltonian are given by
\begin{align}
\{H,C_5\} &= -2H + 4\Lambda\,,  \qquad \quad ~~ \{H,C_6\} = 4H\tilde{\psi} - 8\Lambda\tilde{\psi}\,,\notag \\
\{H,C_7\} &= -4BH + 8\Lambda B\,, \qquad \{H,C_8\} = -4HC + 8\Lambda C\,.
\label{PB_Ci}
\end{align} 
Since the $C_i$ do not depend explicitely on $\tau$, we find immediately that $C_1,C_2,C_3,C_4$ are four
constants of motion besides $H$. Moreover if we define the modified function
${\tilde C}_5\equiv C_5-4\Lambda\tau$ and use the constraint $H=0$, we recover the constant of motion
linked to a scale trasformation $\xi_5$,
\begin{equation}
\frac{\D {\tilde C}_5}{\D \tau} = -2H = 0\,.
\end{equation}
The modification of $C_5$ to ${\tilde C}_5$ is a consequence of the necessity to rescale also
$\Lambda$ in order to maintain invariance under scale transformations.

The only nonvanishing Poisson brackets between the constants of motion read
\begin{align}
\{C_2,C_3\} &= -2C_1\,, \qquad \{C_2,C_4\} = C_3\,, \qquad~~ \{C_3,C_4\} = -C_2\,, \\
\{{\tilde C}_5,C_1\} &= -2C_1\,, \qquad \{{\tilde C}_5,C_2\} = -C_2\,, \qquad \{{\tilde C}_5,C_3\} = -C_3\,.
\end{align}
Among $C_1,C_2,C_3,C_4,{\tilde C}_5$ and the operators composed of them, the maximal set of
commuting first integrals is given by $H,C_1,C_4,C_2^2+C_3^2$, and we fix the values of these first
integrals with four constants $E,v,K_1,K_2$,
\begin{equation}
H = E\,, \quad p_{\tilde\psi} = 4v\,, \quad Bp_C - Cp_B = K_1\,, \quad
(p_B+Cp_{\tilde\psi})^2 + (p_C - Bp_{\tilde\psi})^2 = K_2\,. \label{eq:costanti}
\end{equation}
We want to use these equations to solve the system, so we shall set $E=0$ only at the end of the
integration procedure.

\subsection{Integrability: RN-TN-(A)dS solution}

Using \eqref{eq:costanti}, the Hamiltonian can be rewritten as
\begin{equation}
H = \frac{e^{-\phi}}{2}p_{k}^{2}-\frac{e^{-\phi}}{2k^{2}}p_{\phi}^{2}-\frac{e^{-3\phi}}{8k^{2}}\left(4v\right)^{2}+\frac{e^{-2\phi}}{8k^{2}}\left(K_{2}+16vK_{1}\right)+\hat{V}\,,\label{eq:Hamiltonian-1}
\end{equation}
and thus the electromagnetic and twist part has decoupled from the other fields. In order to solve the
Hamilton-Jacobi equation
\begin{equation}
H\left(k,\phi,\frac{\partial S}{\partial k},\frac{\partial S}{\partial\phi}\right) +
\frac{\partial S}{\partial\tau} = 0\,,
\end{equation}
we use the separation ansatz
\begin{equation}
S = W(k,\phi) - E\tau\,,
\end{equation}
which leads to
\begin{equation}
\frac{e^{-\phi}}2\left(\frac{\partial W}{\partial
k}\right)^{2}-\frac{e^{-\phi}}{2k^{2}}\left(\frac{\partial
W}{\partial\phi}\right)^{2}-\frac{e^{-3\phi}}{8k^{2}}(4v)^{2}+\frac{e^{-2\phi}}{8k^{2}}\left(K_{2}+16vK_{1}\right)+\hat{V} = E\,. \label{reduced-HJ}
\end{equation}
\eqref{reduced-HJ} can be solved by defining the new variables $x=ke^\phi$, $y=e^{-\phi}$ and
applying the Charpit-Lagrange method. The result is 
\begin{eqnarray}
W(x,y) &=& \frac1{6a^2}\sqrt{2ax-v^{2}}\left({\tilde E}(v^2+ax)-6al-12a^2y\right) \nonumber \\
&& +\frac1{8v}(K_2+16v K_1)\text{arccot}\!\left({\frac{v}{\sqrt{2ax-v^{2}}}}\right)\,,
\end{eqnarray}
where $\tilde E\equiv 2\Lambda-E$ and $a$ is an integration constant.
Following the Hamilton-Jacobi technique we can introduce two other constants $\beta_1$, $\beta_2$
according to
\begin{align}
\beta_1 = \frac{\partial S}{\partial{\tilde E}}\,, \qquad \beta_2 = \frac{\partial S}{\partial a}\,.
\end{align}
Using the dynamical constraint $H=0$, they are given by
\begin{equation}
\beta_1 = \frac1{6a^{2}}\sqrt{2ax-v^2}(v^2 + ax) + \tau\,, \label{eq:B1equation}
\end{equation}
\begin{equation}
\beta_2 = \frac{\Lambda (2v^4 - 2av^2x - a^2x^2)}{3a^3\sqrt{2ax-v^2}} + \frac{K_2+16vK_1}{16a
\sqrt{2ax-v^2}} - \frac{lv^2-lax+2a^2xy}{a^2\sqrt{2ax-v^2}}\,. \label{eq:B2equation}
\end{equation}
To simplify the solution, it is convenient to define a new evolution parameter $r$ by
\begin{equation}
\tau = \frac1{\sqrt{2a}}\left(\frac{r^3}3 + r\frac{v^2}{2a}\right)\,. \label{eq:Changecoor-1}
\end{equation}
To solve the two algebraic equations (\ref{eq:B1equation}) and (\ref{eq:B2equation}), we note that it
is possible to set $\beta_1=0$ without loss of generality by shifting $\tau$. Then (\ref{eq:B1equation})
gives
\begin{equation}
x = r^2 + \frac{v^2}{2a}\,. \label{eq: x-equation}
\end{equation}
Plugging this into (\ref{eq:B2equation}) yields
\[
y=\frac1{2a\left(r^2 + \frac{v^2}{2a}\right)}\left[\frac{K_2 + 16vK_1}{16} - \sqrt2\beta_2 a^{3/2}r
+ lr^2 - \frac{lv^2}{2a} - \frac{\Lambda}3\left(r^4 + \frac{3r^2v^2}a - \frac{3v^4}{4a^2}\right)\right]\,.
\]
Using the original expression for $H$ \eqref{eq:Hamiltonian}, the Hamilton equations for the
electromagnetic part become
\begin{align}
\frac{\D p_B}{\D r} &= -\frac{v}{\sqrt{2a}\left(r^2+\frac{v^2}{2a}\right)}(p_C+4vB)\,, \quad~ 
\frac{\D p_C}{\D r} = -\frac{v}{\sqrt{2a}\left(r^2+\frac{v^2}{2a}\right)}(-p_B+4vC)\,, \\
\frac{\D B}{\D r} &= {\frac{1}{4\sqrt{2a}\left(r^2+\frac{v^2}{2a}\right)}}(p_B-4vC)\,, \qquad
\frac{\D C}{\D r} = {\frac{1}{4\sqrt{2a}\left(r^2+\frac{v^2}{2a}\right)}}(p_C+4vB)\,.
\end{align}
Using the gauge freedom generated by $\xi_1$ and $\xi_2$, we can implement 
a boundary condition in such a way that $B$ and $C$ vanish at infinity \cite{Breitenlohner:1987dg}. 
This eliminates two integration constants and the solutions are given by 
\begin{equation}
B = \frac{\beta_3+r\beta_4}{r^2+\frac{v^2}{2a}}\,, \qquad
C = \frac{\sqrt{2a}}v\frac{r\beta_3-\frac{v^2\beta_4}{2a^2}}{r^2 + \frac{v^2}{2a}}\,.
\end{equation}
Finally, the twist potential $\tilde\psi$ can be found by inverting the equation $p_{\tilde{\psi}}=4v$,
which leads to
\begin{equation}
\tilde{\psi}=\int
\D r\left(-\frac{v}{\sqrt{2a}}\frac{e^{-\phi}}{\left(r^2+\frac{v^2}{2a}\right)}-C\frac{\D B}{\D r}+
B\frac{\D C}{\D r}\right)\,.
\end{equation}
The integration procedure is now complete. 
Since the constants defining the solution are not very
illuminating,  we define the new constants
\begin{equation}
m = \frac{\beta_2a^{3/2}}{\sqrt2}\,, \quad n=\frac{v}{\sqrt{2a}}\,, \quad Q=\sqrt{2a}\beta_4\,,
\quad P = -\frac{\sqrt{2a}\beta_3}{n}\,, \quad 2a=m^2+l^2n^2\,,
\end{equation}
which give $K_1=0$ and $K_2=16(P^2+Q^2)$. It turns out that 
the four-dimensional metric and $\text{U}(1)$ gauge field take the form of the RN-TN-(A)dS
solution \cite{Griffiths:2005qp},
\begin{equation}
\D s^2 = -e^{-\phi}(\D t+K_\varphi\D\varphi)^2+ke^\phi\left(\frac{\D r^2}{\Delta}+\D\theta^2
+f_l^2(\theta)\D\varphi^2\right)\,,
\end{equation}
\begin{equation}
A_{\mu}\D x^{\mu} = B\D t+A_{\varphi}\D \varphi\,,
\end{equation}
where 
\begin{align}
\Delta&=l(r^2-n^2)-2mr-\frac{\Lambda}3(r^4+6r^2n^2-3n^4)+P^2+Q^2\,,
\notag \\
k&=\frac{\Delta}{m^2+l^2n^2}\,, \qquad e^{-\phi}=\frac{k}{r^2+n^2}\,, \qquad
K_\varphi = {-4n}\sqrt{m^2+l^2n^2}f_l^2(\theta/2)\,, \\
B &= \frac{Qr-nP}{\sqrt{m^2+l^2n^2}(r^2+n^2)}\,, \qquad
A_{\varphi} = \frac{2f_l^2(\theta/2)\left(P(n^2-r^2)-2nQr\right)}{n^2+r^2}\,.
\notag
\end{align} 
Note that the fields $K_\varphi$ and $A_\varphi$ are obtained from the dualization (\ref{eq:dual1}) and
(\ref{eq:dual2}), that involves 
\[
\tilde{\psi} = \frac{n}{3(m^2+l^2n^2)}\left(\Lambda r+\frac{3lr-3m-4\Lambda n^2 r}{r^2+n^2}\right)\,, \qquad 
\quad C = -\frac{nQ+rP}{\sqrt{m^2+l^2n^2}(r^2+n^2)}\,.
\]
For $P=Q=0$ we recover the results of \cite{Leigh:2014dja}, and thus the integrability properties
described in \cite{Leigh:2014dja} are still valid in the case of nonvanishing electromagnetic charges.
We saw that, even if the cosmological constant reduces the internal symmetry group from
$\text{SU}(2,1)$ to $\text{SU}(2,1)/\text{H}_1$, it hasn't spoiled integrability once we restrict to the
subspace \eqref{eq:subspace}. This condition reduces the infinite number of degrees of freedom
to effectively five.
Only the three nonlinear generators of $\mathfrak{su}(2,1)$ are broken and the remaining commuting
first integrals are enough to decouple the electromagnetic and twist potentials and to integrate the system
in three steps. The general case remains unsolved and is highly linked to the broken affine Kac-Moody
algebra arising after another dimensional reduction \cite{Nicolai:1991tt,Nicolai:2009xxx,Persson:2007xxx}.
The action of $\text{SU}(2,1)/\text{H}_1$ on the fields generates a transformation
on the parameter space, and in particular a scale transformation requires
a rescaling also of $\Lambda$. Unfortunately these surviving symmetries
alone are useless to produce new interesting solutions.

\section{Einstein-$\Lambda$ system}
\label{sec:Einst-Lambda}

In this section, we shall consider the action \eqref{eq:EML action} with vanishing electromagnetic field $F_{\mu\nu}=0$, and assume the existence of an additional Killing vector that commutes with $\partial_t$. For $\Lambda=0$, this system is described by Weyl-Papapetrou formalism which allows us to utilize certain integrability techniques. We see that $\Lambda$ term destroys the reduction to the Weyl-Papapetrou system. In spite of this, a further reduction to $d=1$ with a suitable choice of variables enables us to solve the Einstein-$\Lambda$ system in full generality.

\subsection{Effective field theory in two dimensions}

In general, a solution with an $\mathbb{R}\times\text{SO}(2)$ isometry group cannot
be written in the Lewis form \cite{Gibbons:2008hq}, but this becomes true if the line element admits
a two-dimensional foliation orthogonal to the one in which the action of
$\mathbb{R}\times\text{SO}(2)$ is transitive. With this additional hypothesis we take 
\begin{equation}
\D s^2 = -e^{-\phi}(\D t+K\D\varphi)^2 + e^\phi (e^{2\psi}h_{mn}\D x^m \D x^n +e^{2\chi}\D\varphi^2)\,, 
\label{eq:ansatz assi staz}
\end{equation}
where $t$ and $\varphi$ are Killing coordinates, and the metric depends only on $x^m$ ($m=1,2$). 
Plugging this into \eqref{eq:EML action} (with $F_{\mu\nu}=0$) yields the two-dimensional Lagrangian
\begin{equation}
\mathcal{L}^{(2)}=\sqrt{h}e^{\chi}\left[R^{(2)}+2\partial_{k}\chi\partial^{k}\psi-\frac12(\partial_k\phi
\partial^k\phi-e^{-2(\phi+\chi)}\partial_m K\partial^m K) - 2\Lambda e^{\phi + 2\psi}\right]\,.
\end{equation}
$R^{(2)}$ is the two-dimensional Ricci scalar associated with $h_{mn}$. 
In order to emphasize the nonlinear $\text{SL}(2,\mathbb{R})$ symmetry present for $\Lambda=0$, we
define new fields $\hat\psi$, $\tilde\phi$ by $2\psi=-\phi+\alpha\chi+2\hat{\psi}$ (where $\alpha$
is an arbitrary constant to be specified later), $\phi=\tilde{\phi}-\chi$, which leads to
\begin{equation}
\mathcal{L}^{(2)} = \sqrt{h}e^{\chi}\left[R^{(2)}+2\partial_k\chi\partial^k\hat{\psi}+\left(\frac12+\alpha\right)\partial_k\chi
\partial^k\chi-\langle J_k,J^k\rangle - 2\Lambda e^{\alpha \chi + 2\hat{\psi}}\right]\,,\label{eq:lagra MM}
\end{equation}
where the $\sigma$-model target space is ${\rm SL}(2,\mathbb R)/{\rm SO}(2)$, and thus
\begin{align}
\langle J_k,J^k\rangle = \frac12(\partial_k\tilde{\phi}\partial^k\tilde{\phi}  - e^{-2\tilde{\phi}}
\partial_l K\partial^l K)\,.
\end{align} 
When $\Lambda$ is turned off, we can recover another
$\text{SL}(2,\mathbb{R})$ via dualization in $d=3$ \cite{Breitenlohner:1987dg,Breitenlohner:1998cv,
Breitenlohner:1986um}. Since these two nonlinear representations of
$\text{SL}(2,\mathbb{R})$\footnote{The first $\text{SL}(2,\mathbb{R})$ is called the Matzner-Misner~\cite{Matzner:1967zz} transformation and the
second one the Ehlers transformation~\cite{Ehlers:1957zz}.} are linked by a nonlinear and nonlocal relation, they do not commute. 
This non-commutativity can be used to generate new solutions \cite{Clement:1999bv,Clement:1997tx} and to study the symmetries of the effective field theory (\ref{eq:lagra MM}) (with
$\Lambda=0$) \cite{Breitenlohner:1986um,Nicolai:2001xxx}.

Varying (\ref{eq:lagra MM}) and using the fact that in two dimensions every metric is conformally
flat\footnote{The conformal factor can be absorbed into $\hat{\psi}$, cf.~\eqref{eq:ansatz assi staz}.},
one obtains the equations of motion
\begin{equation}
\Delta e^\chi + 2\Lambda e^{(1+\alpha)\chi+2\hat{\psi}} = 0\,, \label{eq:KG}
\end{equation}
\begin{equation}
\partial_l(e^\chi\partial^l\tilde{\phi} - e^{-2\tilde{\phi}+\chi}K\partial^l K) = 0\,, \qquad
\partial_l(e^{-2\tilde{\phi}+\chi}\partial^l K) = 0\,, \label{eq:sl2 equations}
\end{equation}
\begin{eqnarray}
\partial_{\varrho}^2 e^\chi &=& e^\chi\left[\frac12(\langle J_z,J_z\rangle - \langle J_\varrho,J_\varrho
\rangle) - \partial_z\chi\partial_z\hat{\psi} + \partial_\varrho\chi\partial_\varrho\hat{\psi}\right.
\nonumber \\
&&\quad\left.-\frac12\left(\frac12 + \alpha\right)(\partial_z\chi\partial_z\chi - \partial_\varrho\chi
\partial_\varrho\chi) - \Lambda e^{\alpha\chi+2\hat{\psi}}\right]\,, \label{eq:Einstein 1}
\end{eqnarray}
\begin{equation}
\partial_z\partial_\varrho e^\chi = e^\chi\left[\partial_z\chi\partial_\varrho\hat{\psi} +
\partial_\varrho\chi\partial_z\hat{\psi}+\left(\frac12 + \alpha\right)\partial_\varrho\chi\partial_z\chi -
\langle J_\varrho,J_z\rangle\right]\,,\label{eq:Einstein 2}
\end{equation}
\begin{equation}
\Delta \hat{\psi} - \frac12\left(\frac12 + \alpha\right) \partial_k\chi\partial^k\chi + \frac12 \langle J_k,J^k\rangle -\Lambda \alpha e^{\alpha \chi+2\hat{\psi}} = 0\,, \label{eq:motionpsi}
\end{equation} 
where $\Delta=\partial^2_\varrho+\partial^2_z$. It is interesting to note that for the choice $\alpha=0$
\eqref{eq:KG} becomes the massive Klein-Gordon equation on a curved space with metric
$e^{2\hat\psi}(\D\rho^2+\D z^2)$, and mass $m^2=-2\Lambda$. In the case $\Lambda=0$ it boils
down to $\Delta e^\chi=0$, which allows to take $e^\chi=\rho$ without loss of generality,
but for $\Lambda\neq 0$ $\chi$ does not decouple anymore from the other fields. This explains the failure of the metric to fall into the Weyl-Papapetrou class.  
Notice also that \eqref{eq:motionpsi} follows from
the other equations. To see this, consider
$\partial_z$(\ref{eq:Einstein 1}) - $\partial_\varrho$(\ref{eq:Einstein 2}) and use
(\ref{eq:KG}) and (\ref{eq:sl2 equations}). This leads to
\begin{eqnarray}
\lefteqn{\frac12\partial_z\chi (\partial_z^2 \chi
- \partial_\varrho^2 \chi) + \frac12\partial_z(\partial_\varrho\chi)^2 -\partial_z\hat{\psi}\partial_k \chi \partial^k\chi =} \nonumber \\
&& \alpha\, \partial_k \chi \partial^k \chi \partial_z \chi  - \langle J_z,J^k\rangle \partial_k \chi -\partial_z\chi (\Delta \hat{\psi} -\Lambda \alpha e^{\alpha \chi+2\hat{\psi}})\,. \label{eq:c1}
\end{eqnarray}
Moreover one can rewrite (\ref{eq:Einstein 1})$\cdot\partial_z\chi$ -
(\ref{eq:Einstein 2})$\cdot\partial_\rho\chi$ in the form
\begin{eqnarray}
\lefteqn{\frac12\partial_z\chi (\partial_z^2 \chi
- \partial_\varrho^2 \chi) + \frac12\partial_z(\partial_\varrho\chi)^2   
-\partial_z\hat{\psi}\partial_k \chi \partial^k \chi =} \nonumber \\
&&\left(\frac{\alpha}{2} - \frac{1}{4} \right)\partial_k \chi \partial^k \chi \partial_z \chi
  -\langle J_\varrho,J_z\rangle \partial_\varrho \chi +\frac12 \partial_z\chi (\langle J_\varrho,J_\varrho\rangle - \langle J_z,J_z\rangle)\,. \label{eq:c2}
\end{eqnarray}
Finally, the difference of \eqref{eq:c1} and \eqref{eq:c2} implies (\ref{eq:motionpsi}). 

Note that the $\sigma$-model isometry group $\text{SL}(2,\mathbb{R})$ acts only on $\tilde\phi$ and
$K$. Since the potential in \eqref{eq:lagra MM} is independent of $\tilde\phi$ and $K$, this
$\text{SL}(2,\mathbb{R})$ is a symmetry of the complete Lagrangian \eqref{eq:lagra MM}. However,
one easily checks that, from a four-dimensional point of view, a transformation with this
$\text{SL}(2,\mathbb{R})$ corresponds merely to a diffeomorphism, and can thus not be used to
generate new solutions.

\subsection{Further reduction to $d=1$}

If we make the additional assumption that all the fields depend only on $\varrho$, the two-dimensional
effective field theory (\ref{eq:lagra MM}) boils down to a dynamical system with four degrees of
freedom described by 
\begin{equation}
S= \int \D \varrho e^\chi\left[2\frac{\D \chi}{\D \varrho}\frac{\D \hat{\psi}}{\D \varrho}+\left(\frac12+\alpha
\right)\left(\frac{\D \chi}{\D \varrho}\right)^{\!\!2} - \frac12\left(\frac{\D \tilde{\phi}}{\D\varrho}\right)^{\!\!2} +
\frac{e^{-2\tilde{\phi}}}2\left(\frac{\D K}{\D \varrho}\right)^{\!\!2} - 2\Lambda
e^{\alpha\chi + 2\hat{\psi}}\right]\,, \label{eq:Lagr 1 dim}
\end{equation}
which turns out to be exactly solvable. Introducing the new coordinate $r$ by $e^{-\chi}\D\varrho=\D r$
and defining $V=2\Lambda e^{2\hat{\psi}}$, the system 
(\ref{eq:Lagr 1 dim}) is described by the 
Lagrangian $S=\int L \D r$ as 
\begin{equation}
L = 2\chi'\hat{\psi}' + \left(\frac12 + \alpha\right)\chi'^2 - \frac12({\tilde\phi}'^2 -
e^{-2\tilde{\phi}}K'^2) - e^{(2+\alpha)\chi}V\,, \label{L-1d-r}
\end{equation}
where a prime denotes a derivative w.r.t.~$r$. In what follows, we shall make the choice $\alpha=-2$,
for which $\chi$ becomes cyclic. Then the equations of motion following from \eqref{L-1d-r}
are given by
\begin{equation}
\tilde{\phi}'' - e^{2\tilde{\phi}}A^2 = 0\,, \qquad K' = e^{2\tilde{\phi}}A\,, \label{eq:ODE1}
\end{equation}
\begin{equation}
\hat{\psi}'' + 3\Lambda e^{2\hat{\psi}} = 0\,, \qquad \chi'' + 2\Lambda e^{2\hat{\psi}} = 0\,,
\label{eq:ODE2}
\end{equation}
where $A$ is an integration constant, together with the constraint $H\equiv L+2V=0$, that emerges
from (\ref{eq:Einstein 1}). Eqs.~(\ref{eq:ODE1}) and (\ref{eq:ODE2}) are easily
solved\footnote{The case $A=0$ leads, after a change of coordinates, to $(22.8)$ of \cite{Stephani:2003tm},
and it cannot be recovered smoothly after the integration.}, and one finds that the metric 
in four dimensions is given by
\begin{equation}
\D s^2 = -e^{-\tilde{\phi} +\chi}(\D t + K\D\varphi)^2 + e^{2\hat{\psi}}\D r^2 + e^{2(\hat{\psi}-\chi)}\D z^2 +
e^{\tilde{\phi}+\chi} \D\varphi^2\,,\label{eq:new solution}
\end{equation}
where 
\begin{align}
e^{-\tilde{\phi}}& = \frac{A}{\sqrt{C_1}} \cos[\sqrt{C_1}(r+C_2)]\,, \qquad
e^{2\hat{\psi}} = \frac {C_3}{3\Lambda\cosh^2(\sqrt{C_3}r)}\,,\notag \\
e^\chi &= C_4\frac{e^{r\sqrt{4 C_3 + 3 C_1}/3}}{[\cosh(\sqrt{C_3}r)]^{2/3}}\,, \qquad
K = \frac{\sqrt{C_1}}{A}\tan[\sqrt{C_1} (r+C_2)]\,,
\end{align}
and $C_1$, $C_2$, $C_3$, $C_4$, are integration constants. For generic values of these constants, 
$\partial_t$, $\partial_\varphi$ and $\partial_z$ constitute an exhaustive list of Killing vectors, 
as one can verify by checking the integrability of the Killing equation. 
This solution falls into Petrov type I and the Kretschmann invariant is 
\begin{eqnarray*}
R_{\mu \nu \rho \sigma}R^{\mu \nu \rho \sigma} &=&
\frac{8 \Lambda^2}{3 C_3^{3/2}}[3 C_3^{3/2}\cosh^4(\sqrt{C_3}r) +C_3^{3/2}\sinh^4(\sqrt{C_3}r)\\
&& -2 (C_3 + 3 C_1)\sqrt{4C_3 + 3 C_1}\cosh^3(\sqrt{C_3}r) \sinh(\sqrt{C_3}r)] \,, \label{eq:KrInv}
\end{eqnarray*}
while the Chern-Pontryagin one vanishes. Hence $r\to \infty$ corresponds to a curvature singularity. 
This class of solutions has already been found by Santos
and MacCallum in \cite{MacCallum:1997cn,Santos:1993xxx} and previously by Krasi\'nski
in \cite{Krasinski:1974zza,Krasinski:1900zza,Krasinski:1994zz}, 
in both cases in a different coordinate system. The connection between 
the metric (\ref{eq:new solution}) and those of \cite{MacCallum:1997cn,Santos:1993xxx}  
is easy to find after the identification $ e^{\hat{\psi}(r)} = \Psi({\bf r})$. 
Other explicit forms of the fields defining (\ref{eq:new solution}) can be 
recovered through the combination of analytic continuations, 
diffeomorphisms and nontrivial limits on the constants, 
requiring the final metric to be real.   

An example of a well-known static solution inside the class of (\ref{eq:new solution}) is found  
if we define $A=\sqrt{C_1} \alpha $ and make the choice  
\begin{equation}
C_1 = 0\,, \qquad \alpha = 1\,, \qquad C_3 = 9M^2\,, \qquad C_4 = (\ell^2 M)^{2/3}\,,
\end{equation}
where the parameter $\ell$ is related to the cosmological constant by $\Lambda=-3\ell^{-2}$. Then
the line element boils down to
\begin{equation}
\D s^2 = -e^\chi\D t^2 + e^\chi\D\varphi^2 + e^{2\hat{\psi}}\D r^2 + e^{2(\hat{\psi}-\chi)}\D z^2\,,
\end{equation}
with
\begin{equation}
e^{2\hat{\psi}} = -\frac{M^2\ell^2}{\cosh^2(3Mr)}\,, \qquad
e^\chi = (M\ell^2)^{2/3}\frac{e^{2Mr}}{[\cosh(3Mr)]^{2/3}}\,.
\end{equation}
Introducing the new radial coordinate $R$ by
\begin{equation}
R = (M\ell^2)^{1/3}\frac{e^{Mr}}{[\cosh(3Mr)]^{1/3}}\,, \label{eq:Change coor}
\end{equation}
the solution becomes 
\begin{equation}
\D s^2 = R^2(-\D t^2 + \D\varphi^2) + \left(-\frac{2M}R + \frac{R^2}{\ell^2}\right)^{-1}\!\D R^2 +
\left(-\frac{2M}R + \frac{R^2}{\ell^2}\right)\D z^2\,. \label{eq:Ads planar soliton}
\end{equation}
This is the well-known planar AdS soliton which turns into a planar black hole after the analytic
continuation $t\mapsto iz$, $z\mapsto it$.

Another interesting limit of (\ref{eq:new solution}) is obtained if we choose $C_3=3\Lambda$,
$C_1 =\beta^2$, and, after the limit $\Lambda\to 0$ and the rescaling
$(r,z)\mapsto(\sqrt{C_4}r,\sqrt{C_4^3}z)$, we fix the other parameters as
\begin{equation}
A = \beta = \sqrt{\frac3{C_4}}\,, \qquad C_2 = 0\,,  \qquad C_4 = \frac{1}{k^2} \,.
\end{equation}
Then the metric \eqref{eq:new solution} becomes
\begin{equation}
k^2\D s^2 = -e^r[\cos(\sqrt{3} r) (\D t^2 - \D\varphi^2) + 2\sin(\sqrt{3} r)\D\varphi\D t ] +
e^{-2 r}\D z^2 + \D r^2\,.\label{eq:Petrov}
\end{equation}
This is the Petrov solution \cite{Gibbons:2008hq},  eq.~(12.14), also found 
as the metric induced on a constant $r$ hypersurface of the $d=5$ Einstein-Maxwell-$\Lambda$ 
solution discussed in \cite{Charmousis:2006fx}, eq.~(5.37). It is interesting to note \cite{Stephani:2003tm}
that (\ref{eq:Petrov})
is the only vacuum solution of Einstein's equations admitting 
a simply-transitive four-dimensional maximal group of motions generated by the Killing vectors \cite{Gibbons:2008hq}
\begin{equation}
T=\partial_t\,,\quad Z=\partial_z\,,\quad \Phi=\partial_\varphi\,,
\quad R=\partial_r + z \partial_z + \frac12 (\sqrt3 t - \varphi)\, \partial_\varphi - \frac12(t+\sqrt3\varphi)\,\partial_t\,,
\end{equation}
satisfying
\begin{equation}
[R,T] = \frac12 T- \frac{\sqrt3}2\Phi\,, \qquad [R,\Phi] = \frac12 \Phi+ \frac{\sqrt3}2 T\,, \qquad [R,Z] = -Z\,.
\end{equation} 
Furthermore the determinant of (\ref{eq:Petrov}) is always $-1$, so $(t,\phi,r,z)$ can be promoted to global
coordinates.

It is worthwhile to note that in our approach, which stresses the symmetries of the underlying theory,
a simple system of integrable equations (\ref{eq:ODE1}), (\ref{eq:ODE2}) emerges in a natural way. 
However finding the most general solution to (\ref{eq:KG})-(\ref{eq:Einstein 2}) remains a hard
problem, due to the breaking of the $\text{SL}(2,\mathbb{R})$ Ehlers symmetry, similar to the
Einstein-Maxwell-$\Lambda$ case. As we tried to argue in the introduction, this system of equations
may nevertheless admit a Lax-pair representation, which we expect (if it exists at all) to be
highly nontrivial to find. We hope to come back to this point in a future publication.

\section{Final remarks}
\label{sec:final-rem}

The integrable nature of Einstein's gravity in certain contexts is a useful ingredient for our understanding
of the nonlinear nature of gravitational physics. In this paper, we developed some novel techniques of integrability that allowed us to obtain solutions of the Einstein-Maxwell-$\Lambda$ system. In the first part we derived the complete solution by assuming that every quantity depends only on a single variable and
that the specific form of the base space is $\mathbb R\times \Sigma_l$, by extending the work
of \cite{Leigh:2014dja}. This restriction is strong enough to give a tractable system, yet turns out rich
enough to incorporate a class of gravitational solutions of physical interest. In the second part, we found that under the codimension one assumption, an appropriate choice of variables allows us to obtain the decoupled equations (\ref{eq:ODE1}), (\ref{eq:ODE2}). We expect that these convenient variables would continue to be powerful for further investigations of the case with higher codimension. 

The present work can be generalized into various directions. 
An interesting plausible route is to see if other gravitational theories display integrability properties similar
to the Einstein-Maxwell-$\Lambda$ system in four dimensions.  
As we shall discuss in appendix~\ref{sec:app}, it seems that the higher-dimensional generalization is not straightforward even for the pure Einstein-$\Lambda$ case.  Nevertheless, this failed attempt also gives 
further insight into the integrability properties of Einstein's equations. 

Another possible future work is to extend our formalism to the case where the base space admits the dependence on more than one variable. To this aim, hints that these systems could actually be integrable come from the study of the Pleba\'nski-Demia\'nski solution \cite{Plebanski:1976gy}, which is the most general known Petrov-type D solution of the four-dimensional Einstein-Maxwell equations with $\Lambda$. It possesses a very high degree of symmetry and contains a lot of subcases and parameters of physical interest, so one may think of being able to generate it from a given seed solution. Also its relation to 
supersymmetry \cite{Klemm:2013eca} may be a tractable way to understand the integrable nature of
Einstein's gravity in the presence of a cosmological constant. We hope to come back to these points in
future work.

\section*{Acknowledgements}

This work was supported in part by INFN and JSPS.

\appendix

\section{On the integrability of the higher-dimensional Einstein-$\Lambda$ system}
\label{sec:app}

In this appendix we discuss if the integrability properties in four dimensions 
can be extended to higher dimensions. 

Let us consider the $D$-dimensional spacetime 
($M_D, g_D$) represented by Einstein's gravity with a cosmological constant,
\begin{align}
S_D=\frac 1{2\kappa_D^2} \int \D ^D x\sqrt{-g^{(D)}} 
\left(R^{(D)}-2 \Lambda \right)\,.
\end{align}
Suppose that ($M_D, g_D$) admits $n$ Killing vectors which mutually commute.
The metric can be put into the form 
\begin{align}
 \D s^2 _{D} = \hat g_{\mu\nu }(x) \D x^\mu \D x^\nu 
+\ti M_{mn}(x)(\D y^m +2K^{(m)}_{\mu }(x)\D x^\mu )
(\D y^n +2K^{(n)}_{\nu }(x)\D x^\nu ) \,.
\label{KK_on_ntorus}
\end{align}
Here $\hat g_{\mu\nu}$ is the metric of the external 
$d\equiv D-n$ dimensional space, $\ti M_{mn}$ is the internal metric
and $K^{(m)}_\mu$ are the Kaluza-Klein gauge fields. 
Due to the isometries, we can reduce the $D$-dimensional gravity system down to
$d$ dimensions. Denoting the Kaluza-Klein field strength as 
$K^{(m)}_{\mu\nu}=\partial_\mu K^{(m)}_\nu-\partial_\nu K^{(m)}_\mu$, 
it is straightforward to show that 
the Ricci scalar is decomposed as
\begin{align}
R^{(D)} =&\hat R-\ti M_{mn}
K^{(m)}_{\mu\nu}K^{(n)\mu\nu}
-\ti M^{mn}\hat \nabla^2 \ti M_{mn} \notag \\
&-\frac 14 (\ti M^{mn}\hat \nabla
 \ti M_{mn})^2 -\frac 34 \hat \nabla ^\mu\ti M^{mn}\hat \nabla _\mu \ti M_{mn}\,,
\end{align}
where $\hat \nabla_\mu$ is the linear connection compatible with $\hat g_{\mu\nu}$. 
Let us perform the conformal transformation
\begin{align}
 \hat g_{\mu\nu }=ke^{\alpha \phi } g_{\mu\nu }\,, \qquad \ti
 M_{mn}=e^{\beta\phi }M_{mn} \,, \qquad 
{\rm det}(M)=-s \,,
\end{align}
where $s=\pm 1$ and the additional scalar field $k$ has been introduced as in the body of text. To achieve the Einstein frame for $k=1$,  we choose 
\begin{align}
 \alpha =\sqrt{\frac{2n}{(d+n-2)(d-2)}} \,, \qquad 
\beta =-\frac{(d-2)\alpha }{n} 
\label{KK_alphabeta}
\end{align}
One can then verify that the $d$-dimensional equations of motion 
can be derived from the action
\begin{align}
S_d =\frac{1}{2\kappa_d^2} \int \D ^d x \sqrt{|g|} k^{(d-2)/2} &\left[R- \frac12 (\nabla\phi)^2 
+\frac 14 {\rm Tr}(\nabla M \nabla M^{-1})
 +\frac 14 
 (d-1)(d-2)\frac{(\nabla k)^2}{k^2}
 \right.\nonumber \\ & \left.  
 -k ^{-1} e^{(\beta-\alpha
 )\phi}M_{mn}K^{(m)}_{\mu\nu } K^{(n)\mu\nu } 
-2\Lambda ke^{\alpha\phi}\right]\,. 
\label{daction_KK_on_ntorus}
\end{align}
Hence $M_{mn}$ is the matrix corresponding to an ${\rm SL}(n,\mathbb R)/{\rm SO}(n)$ 
nonlinear sigma model.

In the following, 
we shall focus on the $d=3$ case. This is a generalization of the 
vacuum \cite{Maison:1979kx} and electrovac \cite{Ida:2003wv} analyses.
The Maxwell equations can be solved to give local twist potentials 
$\psi_{(m)}$ such that 
\begin{align}
k^{-1/2} e^{(\beta-\alpha)\phi} M_{mn}K^{(n)\mu\nu} =\frac 12 \epsilon^{\mu\nu\rho} \nabla_\rho \psi_{(m)} \,, 
\end{align}
where $\epsilon_{\mu\nu\rho}$ is the volume element compatible with 
the external metric $g_{\mu\nu}$.  
It therefore turns out that 
the 3-dimensional Einstein- and matter field equations can be derived from the action 
\begin{align}
S_3 =\frac {1}{2\kappa_3^2} 
\int \D ^3 x \sqrt{|g|} k^{1/2} \left[
R
+\frac 14 {\rm Tr}(\nabla\mathbb M \nabla \mathbb M^{-1})
+\frac 1{2k^2}(\nabla k)^2 -2 \Lambda ke^{\alpha \phi} 
\right]\,,
\end{align}
where 
\begin{align}
\label{boldM}
 \mathbb M&= \left(
\begin{array}{cc}
({\rm det}\ti M )^{-1}
 & - ({\rm det}\ti M )^{-1} \psi_{(m)} \\
- ({\rm det}\ti M )^{-1}\psi_{(n)} &~ \ti M_{mn}+ ({\rm det}\ti M )^{-1}
\psi_{(m)} \psi_{(n)}  
\end{array}
\right)\,, \quad {\rm det}\mathbb M=1 \,, \quad \mathbb M
=\mathbb M^T\,,
\end{align}
with inverse 
\begin{align}
\mathbb M^{-1}&=\left(
\begin{array}{cc}
{\rm det}\ti M+\ti M^{pq}\psi_{(p)} \psi_{(p)} ~
 & \ti M^{mp} \psi_{(p)} \\
\ti M^{np}\psi_{(p)} & 
\ti M^{mn}
\end{array}
\right) \,.
\end{align}
It turns out that the coset symmetry is now enhanced to 
${\rm SL}(D-2)/{\rm SO}(D-2)$. 

Let us focus on the $D=5$ case in what follows. 
Assuming two commuting Killing vectors 
$\xi_{(1)}=\partial/\partial t$ and $\xi_{(2)}=\partial/\partial \psi$, 
we parametrize  
\begin{align}
\label{}
\ti M_{mn}\D y^m \D y^n=
-f (\D t+\omega\D \psi)^2+(fh)^{-1} \D \psi^2 \,, \qquad 
h=e^{2\phi/\sqrt 3} \,. 
\end{align}
It follows that 
the matrix $\mathbb M$ in (\ref{boldM}) can be expressed by 
\begin{align}
\label{}
\mathbb M=\left(\begin{array}{ccc}
-h & h\psi_{(1)} & h \psi_{(2)} \\
h\psi_{(1)}  & -f-h \psi_{(1)}^2 & -f\omega -h\psi_{(1)}\psi_{(2)} \\
h\psi_{(2)}  & ~-f\omega -h\psi_{(1)}\psi_{(2)}  ~& 
-h \psi_{(2)}^2-f \omega^2+(fh)^{-1} 
\end{array}\right)\,.
\end{align}
This coincides with the parametrization  ($q_1,q_2,q_3, p_1,p_2,p_3$)
of ${\rm SL}(3,\mathbb R)/{\rm SO}(3) $ given in \cite{Galtsov:1995mb}  by 
the identification
\begin{align}
\label{}
q_1\mapsto -h \,, \quad q_2 \mapsto -f \,, \quad 
p_1\mapsto -\psi_{(1)} \,, \quad 
p_2\mapsto  \omega \,, \quad 
p_3\mapsto -\psi_{(2)} \,.  
\end{align}
The Killing vectors $\hat X_i$ ($i=1,...,8$) 
of the target space $\D s_\Phi^2 =-(1/4){\rm Tr}(\D \mathbb M\D \mathbb M^{-1})$
are given by (5.24) of \cite{Galtsov:1995mb}\footnote{$\hat X_6$ seems to have a typo in
ref.~\cite{Galtsov:1995mb}  and it should be modified to $\hat X_6=2q_2p_2 \partial_{q_2}+p_3 \partial_{p_1}+(q_1^{-1}{q_2^{-2}}-p_2^2)\partial_{p_2}$.}.

For $k=1$, the potential depends only on $\phi$. The Killing vectors which keeps the potential invariant are given by
\begin{align}
\label{inv_Killing}
\hat X_-\equiv \hat X_1-2\hat X_2 \,, \qquad \hat X_3\,, \qquad \hat X_4 \,, \qquad 
\hat X_5 \,, \qquad \hat X_6 \,.
\end{align}
The remaining three generators ($\hat X_+\equiv \hat X_2-\hat X_1, \hat X_7, \hat X_8$)
do not leave the potential invariant, and form a ${\rm Sim}(1)\times \mathbb R$ algebra 
corresponding to the Bianchi III universe, 
\begin{align}
\label{}
[\hat X_+,\hat X_7]=\hat X_7\,, \qquad [\hat X_+, \hat X_8]=[\hat X_7,\hat X_8]=0 \,. 
\end{align}
We make a codimension one ansatz and require that the 
base space metric has an ${\rm SO}(3)$ symmetry. 
Then, the base space reads 
\begin{align}
\D s_{\text B}^2 =\D x^2+\sigma_1^2 +\sigma_2^2\,, 
\end{align}
where $\sigma_i$ are ${\rm SU}(2)$ invariant forms,
\begin{align}
\sigma_1 =-\sin\psi \D \theta +\cos\psi \sin\theta \D \varphi \,, \quad 
\sigma_2 =\cos\psi\D \theta +\sin\psi \sin\theta \D \varphi \,, \quad 
\sigma_3 =\D \psi + \cos\theta \D \varphi \,. 
\notag
\end{align}
Every quantity is dependent only on a single variable $x$. This class of metrics 
includes the five-dimensional Myers-Perry black hole~\cite{Myers:1986un} with equal angular momenta. 
Defining 
\begin{align}
\label{}
 k^{3/2} h\D x= \D \tau \,, \qquad \hat V=2\Lambda -2 k^{-1} 
h^{-1}\,,
\end{align}
the one-dimensional Lagrangian $S=\int L \D \tau$ reads
\begin{align}
\label{}
L=\frac 12 h k '{}^2+\frac 14 k^2 h {\rm Tr}[(\mathbb M)'(\mathbb M^{-1})']-\hat V \,,
\end{align}
where the prime denotes differentiation w.r.t.~$\tau$. 
The Hamiltonian boils down to
\begin{align}
\label{h5}
H=&\frac 12 \Bigl[
h^{-1}p_k^2+f^{-2}h^{-2}k^{-2}p_\omega^2 +f^{-1} h^{-3}k^{-2} p_{\psi_2}^2 
- f h^{-2}k^{-2} (p_{\psi_1}+\omega p_{\psi_2})^2
\notag \\
&-\frac 4{3k^2} \left(f^2 h^{-1} p_f^2-f p_fp_h +h p_h^2\right) \Bigr]+\hat V \,.
\end{align}
The trace of Einstein's equations gives the constraint $H=0$.
For $\varphi^I=\{f,\omega, h, \psi_{(1)}, \psi_{(2)}\}$, 
one can verify that 
$C_3, C_4, C_5, C_6$ are obvious first integrals and other nonvanishing 
Poisson brackets with the Hamiltonian are 
\begin{align}
\label{}
\{C_1 , H\} &=-\frac 43 H+\frac 83\Lambda \,, \qquad \qquad ~~~
\{C_2, H\} =-\frac 23 H+\frac 43 \Lambda\,,  \\
\{C_7, H\}&= -2\psi_{(1)} H+4\Lambda \psi_{(1)} \,, \qquad 
\{C_8, H\}=-2 \psi_{(2)} H +4\Lambda \psi_{(2)} \,.   
\end{align}
Hence $C_-\equiv C_1-2C_2$ is another first integral. 
The nonvanishing Poisson bracket among these first integrals read
\begin{align}
\label{}
 & \{C_3, C_4\}  =-C_5 \,, \qquad 
    \{C_4, C_6\} =C_- \,, \qquad 
      \{C_5, C_6\}=C_3 \,, \qquad   \{C_-, C_3\} = C_3\,,
      \notag    \\
    &\{C_- , C_4\}=-2 C_4\,, \qquad 
      \{C_- , C_5\}    =-C_5 \,, \qquad 
        \{C_- , C_6\}    =2 C_6 \,.  
\end{align}
Thus $C_4, C_6, C_-$ generate an ${\rm SL}(2,\mathbb R)$ subalgebra and
its quadratic Casimir $C_-^2 -4 C_6 C_4 $ is another obvious first integral.
Clearly, this is not true for the ${\rm SL}(3,\mathbb R)$  quadratic Casimir
$ \frac1{3} (C_1^2 - C_1 C_2 + C_2^2 - C_4 C_6 - C_3 C_7 - C_5 C_8) $
which involves also generators of the broken symmetries.

We found that three is the maximal number of commuting first integrals in involution with $H$
and there are many sets of this type that can be built from $C_-,C_3, C_4, C_5, C_6$ and their compositions. 
In order to decouple some of the fields in the Hamilton-Jacobi equation associated to (\ref{h5}), the most
intriguing set seems to be composed of $H, C_3, C_5$ and  the cubic invariant
$\ma C^{(3)}=C_3^2C_4+C_5^2C_6-C_3C_5 C_-$.  
Unfortunately, we were unable to find a sufficient number of commuting first integrals to carry out
the integration procedure as in the four-dimensional Einstein-Maxwell system. 
One plausible reason for this is that 
the target space ${\rm SL}(3, \mathbb R)/{\rm SO}(3)$ is five-dimensional admitting 
eight Killing vectors, while in the $D=4$ Einstein-Maxwell case the target space (\ref{target}) is
four-dimensional with eight Killing vectors. Obviously, the former is less symmetric. 
It would be interesting to see if this system displays chaotic behavior. We shall leave this point for
future investigation.

\end{document}